# Classification of super domains and super domain walls in permalloy antidot lattices

X.K. Hu,[1,2,*] S. Sievers,[1] A. Müller,[1] V. Janke,[1] and H.W. Schumacher[1]

[1]Physikalisch-Technische Bundesanstalt, Bundesallee 100, D-38116 Braunschweig, Germany

[2]College of Materials Science and Engineering, China Jiliang University, Hangzhou 310018, China.

Corresponding author:

Xiukun Hu

AG 2.52 Nanomagnetismus

Physikalisch-Technische Bundesanstalt

Bundesallee 100, 38116

Braunschweig, Germany

Tel: +49(0)531 592 1412

E-mail: xiukunhu@hotmail.com

**ABSTRACT**

We study the remanent domain configurations of rectangular permalloy antidot lattices over a range of lattice parameters. The influence of antidot diameter, antidot spacing, and the aspect ratio of the lattice on the remanent domain configuration are investigated by magnetic force microscopy and supported by micromagnetic simulations. In the remanent state, areas of cells with the same orientation of average magnetization form magnetic super domains separated by super domain walls (SDWs). Two types of SDWs are identified. The first type is characterized by low stray field energy, is linear, and expands over many lattice constants. In contrast the second type shows high stray field energy and is situated at kinks of low energy SDWs. Its width can vary from a minimum of two lattice cells up to several lattice constants, depending on the lattice parameters. The occurrence and structure of these two types of SDWs as function of lattice parameters are classified and discussed in terms of the interplay of stray field and exchange energy.

## I. INTRODUCTION

Patterned ferromagnetic films containing antidot lattices, which allow engineering of the magnetic properties of magnetic thin films,[1,2] are of significant interest. Antidot lattices with different filling factors $f$ (i.e. a ratio of diameter $d$ and periodicity $s$) are considered attractive candidates for a variety of potential applications. Although antidot lattices with a high filling factor (i.e. $s \leq 2d$) are considered, for example high-density storage media,[2-4] antidot lattices with a low filling factor ($s > 2d$) are considered magnonic crystals required for building magnetism-based logic devices that use spin waves rather than electrons for information processing.[5] Much of the literature has focused on novel domain configurations[6-11], additional magnetic anisotropy[1,3,8], magnetization reversal[4,7] and magnetic properties[12-14] in the antidot lattices, with different antidot shapes and sizes over the past years.

The domain structure that develops in response to the introduction of antidot lattices into magnetic films is characterized by a long range ordered magnetization pattern with the periodicity of the lattice. In rectangular antidot lattices the domain structure in a lattice cell can be described roughly as shown in Fig. 1(a).[3,6,8,10,11] The structure is characterized by a large central domain with an average magnetization that points in one of the four diagonal orientations and thus by the corresponding average magnetization vector $m_{av}$. Areas of the cells with the same orientation of $m_{av}$ could then be referred to as magnetic super domains (SDs) separated by super domain walls (SDWs). The domain structure in the antidot lattices as well as the structure and the mobility of the domains determines the magnetic properties of the antidot lattice. The magnetization reversal, for example, could be mediated by a motion of SDWs. Among the many studies of antidot lattices,[1-14] only a few investigations considered the structure and properties of SDW.[7,10,11] Mengotti *et al*. characterized the domain structure in cobalt antidot lattices with large antidot diameters $d \geq 200$ nm and antidot size $\geq$ antidot separation.[7] Heydermann *et al*. reported on chains of domains in cobalt antidot square arrays and assumed that the chains of domains resulted from the diagonally joining antidots.[11] However, a comprehensive study of the structure of SDWs over a large range of lattice parameters has not been published until today.

In this paper we investigate the remanent domain configurations of lattices of circular antidots in a 30 nm thick permalloy film with *small* antidot diameters ($d$=60, 80, and 100 nm) and a *low* filling factor ($\leq 0.6$) in permalloy films. A detailed systematic characterization of the influence of antidot diameter, antidot spacing and aspect ratio of the lattice is carried out with magnetic force microscopy (MFM) supported by micromagnetic simulation. The magnetization structures of the cells and the complete lattices were analyzed. Before the MFM characterization, the films were saturated by applying a magnetic field parallel to the lattice rows. In the remanent state two types of SDWs were found in the lattices: low stray field energy (LE) and high stray field energy (HE) SDWs. LE-SDWs are linear and expand over many lattice constants. HE-SDWs occupy a few lattice cells and are situated at kinks of LE-SDWs. Because of a high divergence of stray field in HE-SDWs, strong interaction with the magnetic tip of the MFM is present, which can lead to the generation, displacement, or deletion of extended HE-SDWs during the scanning process. The occurrence and structure of HE-SDWs are summarized in phase diagrams as a function of the lattice parameters. The appearance of different phases, the micromagnetic structure, and the mobility of the HE-SDWs are discussed in terms of an interplay of stray field and exchange energy.

## II. EXPERIMENTAL DETAILS

A 30nm thick permalloy film containing magnetic antidot lattices with circular antidot diameters ($d$=60, 80, and 100 nm) and spacings $S_x$(100-350 nm), $S_y$(50-1100 nm) [Fig.1(a)] was fabricated by e-beam lithography and lift-off process. The area of every single lattice with the given lattice parameters was 10×10 $\mu m^2$. The domain configurations were investigated by a Veeco magnetic force microscope NanoScope IIIa with a high resolution MFM tip (Team Nanotec MFM probe with 25 nm Co alloy coating, 3 N/m spring constant and 75 kHz resonance frequency). Micromagnetic simulations of the different lattice domain structures were performed using Landau-Lifshitz-Gilbert (LLG) micromagnetic simulator.[15] The computational size was 4200×4200×30 $nm^3$, with an antidot array of 2800×3000 $nm^2$. The simulation cell size was 10×10×10 $nm^3$. Typical intrinsic parameters of permalloy were initialized in the simulations as follows: saturation magnetization $\mu_0 M_S$=1.005 T, exchange stiffness $A$=1.05×10$^{-11}$ J/m, and uniaxial anisotropy constant $K_u$=100 J/m$^3$ with the orientation of the intrinsic uniaxial easy axis of the unpatterned film in a diagonal direction along the antidot lattice.

The MFM images, as well as the simulations, were performed in a remanent state after applying an external saturating field of 1 T parallel to the rows ($-x$ direction). A large number of SDs and SDWs are thereby generated. After removing the external field, the average magnetization $m_{av}$ of the individual cell can relax into either of the two basic directions adjacent to $-x$ direction-i.e. average magnetization vector $m_{av}$ points in a [-1,1] (*up*) or [-1,-1] (*down*) direction-as shown in Fig. 1(b). Areas of cells with parallel average magnetization are separated from neighboring areas with different magnetization directions by SDWs. The SDWs can be subdivided into sections along $x$- or $y$-axis. SDWs along the $x$-axis represent interfaces between

*up*- and *down*-oriented rows, i.e. either head-to-head or tail-to-tail configurations of the average magnetization vector, as shown in Fig. 1(c) where a tail-to-tail configuration is shown. Both configurations exhibit high energy, because they create a high divergence of stray field and high exchange energy density, and hereafter are referred to as HE-SDW. Correspondingly SDWs along *y*-axis show head-to-tail [Fig. 1(d)] or tail-to-head configurations with low stray field divergence and lower exchange energy terms and therefore are referred to as LE-SDW. Consequently, after decay of the saturating field in the −*x* direction, we expect (1) a relaxation of the magnetization within the cells into one of two preferred orientations, the [-1,1] or the [-1,-1] direction and (2) a nucleation and growth of SDs driven by the high energy of the HE-SDW.

## III. INFLUENCE OF THE ANTIDOT DIAMETER

In this section, we classify SDs and SDWs in MFM images and discuss the influence of antidot diameter *d* on the remanent spin configurations of SDs and SDWs. We show the tendency of an increased width of HE-SDW with decreasing *d*, which is supported by micromagnetic simulations.

Figure 2 shows a typical MFM image of the lattice with *d*=80 nm, $S_x$=350 nm, and $S_y$=300 nm in the remanent state. As described earlier and in accordance with the results reported in most references,[1,4,7,8,10] all cells in the lattice show the basic domain structure consisting of a large central domain with an average magnetization vector pointing diagonally *up* or *down*. The preferable overall orientation to the left is determined by the orientation of the saturating field. Cells with the same orientation of $m_{av}$ prefer to form columnar SDs in the *y* direction, expanding over the antidot lattice, and thereby avoiding HE-SDW. These SDs span one or several columns, building one-dimensional (1D) or two-dimensional (2D) SDs as marked in Fig. 2. They are separated by LE-SDW, as indicated by the dashed lines. Also, HE-SDWs, marked by the dotted rectangular frames, are observed in Fig. 2. HE-SDWs are present at the intersections of two SDs in the same column. At these intersections, the average magnetization vectors above and below this SDW are orientated head to head or tail to tail. These domain walls thus only occur at kinks of LE-SDWs. In the lattice under discussion, the *widths* of the HE-SDWs (i.e. their extension in the $y$ direction) can cover several lattice cells.

The different appearances of the two types of SDWs in MFM images can be better understood by qualitatively analysis of their microscopic spin configurations. Therefore, micromagnetic simulations were performed using an LLG micromagnetic simulator on a permalloy antidot lattice with $d$=80 nm, $S_x$=350 nm and $S_y$=300 nm. Typical results are shown in Fig. 3(a). The domain structure consists of the basic diagonal domains and both types of SDWs, as discussed previously. Columnar SDs and a HE-SDW marked by a rectangular frame occur in the image. The figures show the details of the spin configurations within a SD [Fig. 3(b)], along a LE-SDW [Fig. 3(c)], and along a HE-SDW [Fig. 3(d)]. In Fig. 3(c), the central antidot column corresponds to the center of the LE-SDW which is marked by a dashed line. Surrounding these antidots, magnetic spins diverge at the upper right, align along the antidot boundary, make a half circle, and then converge at the upper left. In this spin configuration, no strong out-of-plane stray field is generated. The energy density of LE-SDW calculated by micromagnetic simulations is $3.3 \times 10^3$ J/m$^3$, only 3% higher than that of cells within a SD ($3.2 \times 10^3$ J/m$^3$) as shown in Fig. 3(b). The driving force for a reduction of the LE-SDW density is therefore small, and a lattice with LE-SDW is metastable. For the HE-SDW shown in Fig. 3(d), a different situation occurs. The HE-SDW which is marked by a rectangular frame, is situated between the central antidot column and its neighbouring antidot column on the right hand side. The spins in HE-SDW align along $-x$. The magnetization is orientated perpendicular to the antidot boundary between two antidots in the $x$ direction. A strong out-of-plane stray field is generated. Therefore, the HE-SDW is well visible as a dark or bright region in the MFM image, in accordance with experimental results. The energy density of HE-SDW, as shown in Fig. 3(d), is $6.0 \times 10^3$ J/m$^3$, almost twice as high as that of cells within the SD.

MFM images of one typical lattice with the smaller antidot diameter $d$=60 nm are shown in Figs. 4(a) and (b). The lattice parameters are $d$=60 nm, $S_x$=350 nm, and $S_y$=300 nm. Also, this lattice shows 2D and 1D SDs and LE-SDWs, as well as HE-SDWs, as discussed previously. However, here the HE-SDWs extend over a significantly larger number of lattice cells than in the case of $d$=80 nm. The leftmost HE-SDW in Fig. 4(a) expands over > 15 lattice cells; the other two HE-SDWs in the image occupy 13 and 7 cells, respectively. The general domain structure in this HE-SDW is similar to that in the lattice with larger antidot diameter $d$=80 nm but can show more complex substructures. Fig. 4(b) shows a magnified detail of one HE-SDW. In one cell of this HE-SDW, the horizontal spin configuration is partially split into two diagonal domains, as marked by the two scissored arrows. In contrast, the antidot lattice with $d$=100 nm, $S_x$=300 nm, and $S_y$=300 nm does not show any extended HE-SDW. Figure 4(c) shows the type of SD structure found to be characteristic for this antidot lattice. Here 1D SDs are embedded in a large 2D SD and end with pairs of HE-SDWs. Different contrasts at both ends are related to inverted spin configurations, as shown in Fig. 4(d). In this lattice, the width of the HE-SDW is only two cells, corresponding to the minimum cell size for the occurrence of a HE-SDW. The experimentally observed dependence of HE-SDWs on antidot size was confirmed by micromagnetic simulations. Lattices with $S_x$=350 nm, $S_y$=300 nm have been modeled using different antidot diameters. With a decreasing antidot diameter from 120 nm to 80 nm, the simulations show an increasing width of the HE-SDW from three to four lattice cells. However, despite the confirmation of the general tendency toward increased width of the HE-SDWs with decreasing antidot diameter, very long HE-SDWs extending over more than four cells, as shown in Fig. 4(a), have not been observed in the simulations.

## IV. INFLUENCE OF THE LATTICE SPACING

As explained earlier, antidot diameter has a strong influence on the occurrence and structure of HE-SDWs. In the following we show that these properties also depend on antidot lattice spacing. In our work, a large number of magnetic antidot lattices with $d$=60, 80, and 100 nm, $S_x$=100-350 nm and $S_y$=50-1100 nm have been characterized and their typical SDW configurations have been analyzed. Occurrences of different types of SDWs as a function of lattice parameters are compared in Fig. 5. LE-SDWs appear for all lattice parameters under investigated. The regions in which additional HE-SDWs occur are shaded. Open circles mark lattices where no HE-SDWs are found and only LE-SDWs occur. Half-filled (red) circles mark lattices where HE-SDWs with only a narrow width (two lattice cells) appear. Filled (blue) circles mark lattices where extended HE-SDWs are found. Here, the number of HE-SDW decreases with decreasing lattice spacing. Two typical MFM images of the lattice with extended (upper right) and narrow (lower right) HE-SDWs are shown to the right side of Fig. 5. The phase diagrams show that, with increasing antidot diameter, the left border of region of occurrence of HE-SDW moves towards the right (i.e. towards the larger lattice spacing). In addition, extended HE-SDW (filled circles) move to the region with the larger spacings $S_x$ and $S_y$ (i.e. to the upper right) and stable HE-SDW (half-filled circles) are dominant with increasing $d$.

The influence of the lattice parameters on the occurrence and structure of HE-SDWs can be discussed with respect to Fig. 6. The occurrence and structure of HE-SDWs are a function of antidot diameter and lattice spacings. They can be qualitatively explained as consequence of an interplay between stray field and exchange energy. In the center part of a HE-SDW, magnetic spins align in the $x$ direction, separating the *up* and *down* magnetization directions of the adjacent cells. The sections between two antidots exhibit a high energy density, because the high stray field diverges from the edge of the antidots. However, in the middle area of one cell, the parallel orientation of magnetization on both sides reduces the exchange energy density. For lattices with small antidots as shown in Fig. 6(a), the low energy density section is comparably wide with respect to the high energy density section. Hence, an expansion of the HE-SDW reduces the exchange energy contribution, favoring an extended HE-SDW. In contrast, the high energy density section becomes wider than the lower one when increasing $d$, as shown in the left part of Fig. 6(b). Thus, the additional stray field energy is no longer compensated for by a gain of exchange energy, and the total energy in each cell of the HE-SDW increases, favoring narrow HE-SDWs [right part of Fig. 6(b)], as confirmed by the preceding micromagnetic simulations. Correspondingly, the high stray field energy density section increases with decreasing spacing for the lattices with same $d$, thereby increasing the total energy in one cell. Therefore HE-SDWs, become narrower to decrease the total system energy with decreasing spacings.

## V. STABILITY OF HIGH-ENERGY SUPER DOMAIN WALLS

As discussed previously, HE-SDWs show a high divergence of stray field and therefore strong interaction with the magnetic stray field of the MFM tip. For the lattices with lattice parameters that allow for extended HE-SDWs (shaded areas with filled circles in the phase diagrams of Fig. 5), this mediates an interaction activated motion of extended HE-SDWs or of their edge regions. During the scanning process, the width of HE-SDWs may shrink or extend and HE-SDWs can be deleted or nucleated. The extended HE-SDWs in these lattices were changed with every new image captured in succession. The lattice with $d$=80 nm, $S_x$=350 nm, and $S_y$=300 nm is a typical example. Fig. 7 (a) and (b) show two consecutive MFM images of this lattice. In these pictures some HE-SDWs that disappeared (1), whose widths were increased (2), and whose widths shrank (3) are marked with arrows. The schematic map in Fig 7(c) compares the domain structures of the section marked by dotted rectangular in Fig. 7(b) from scan to scan during five consecutive MFM images (indicated by different colors online). Fig. 7(d) shows a histogram compiling the changes from scan to scan over 26 successive MFM scans. The number of occurrence of the changes in width ($\Delta$W) of extended HE-SDWs in a unit of lattice cells is plotted. Unchanged HE-SDWs are not included in the histogram. The data can be approximated by a Gaussian distribution. Changes of $\Delta$W from -23 to 25 cells are observed. The probability that $\Delta$W changes in the range of ±5 lattice cells is 60%. These images show a high mobility of the HE-SDW and high degree of the MFM tip induced fluctuations of SDWs during the MFM scans for the given lattice parameters. Note that, some lattice cells marked with circled + in Fig. 7(c) were occupied with sections of HE-SDW during all scans, pointing to a pinning of the domain wall structure at these sites. The mechanisms of such pinning are not yet understood.

However, a simple antidot lattice defect resulting from missing antidots can be excluded by comparison of the MFM images to the simultaneously recorded topographic images.

The stability of the HE-SDW can be illustrated as follows. As discussed previously, for lattices with small antidots as shown in [Fig. 6(a)], the total energy in one cell of an extended HE-SDW is relatively low, because the reduction of exchange energy in the wide low energy density section can compensate for by the increase of stray field energy in the narrow high energy density section. The energy barrier for the expansion or contraction of one lattice cell of a HE-SDW is thus low. In contrast, larger antidots cause an increase of total energy in one cell because of a wide high stray field energy density section. The switching of the magnetization is then blocked by the high energy barrier. This results in a pinning effect that makes a narrow HE-SDW in the lattice with large antidots more stable than extended ones.

## VI. CONCLUSIONS

We have investigated the remanent domain configurations of circular antidot lattices in a 30 nm-thick permalloy film with small antidot diameters ($d$=60, 80, and 100 nm) over a wide range of lattice spacings ($S_x$=100-350 nm and $S_y$=50-1100 nm). The characterization of the influence of antidot diameter, antidot spacing and aspect ratio of the lattice was carried out with MFM supported by micromagnetic simulations. In the remanent state, two types of super domain walls were found in the lattices: one is characterized by low stray field energy, is linear and expands over many lattice constants; the other is characterized by high stray field energy with a width that depends on the lattice parameters. Narrow HE-SDWs with the minimum width of two lattice cells are stable, whereas extended ones are mobile. The occurrence and structure of two types of SDWs as function of lattice parameters were classified and discussed in terms of an interplay of stray field and exchange energy.

**ACKNOELEDGMENTS**

The research was performed within a Euramet joint research project and receives funding from the European community's Seventh Framework Programme, EEA-NET Plus, under IMERA-Plus Project-Grant Agreement No. 217257.

**Figure Captions:**

FIG. 1 (a) Schematic diagram of the permalloy antidot lattice and the domain distributions in one cell. (b) Two possible orientations of $m_{av}$ in experimental conditions. (c) HE-SDW in the $x$ direction. (d) LE-SDW in the $y$ direction.

FIG. 2 Remanent state MFM images of the lattice: $d$=80 nm, $S_x$=350 nm and $S_y$=300 nm. Solid circles indicate the positions of antidots. Dashed lines and dotted rectangular frames indicate the LE-SDWs (L) and HE-SDWs (H), respectively.

FIG. 3 The spin configurations of the lattice with $d$=80 nm, $S_x$=350 nm, and $S_y$=300 nm (a) and close-ups of SD (b), LE-SDW (c) and HE-SDW (d) obtained numerically. HE-SDWs are indicated by rectangular frames in (a) and (d). LE-SDW is indicated by the dashed line in (c).

FIG. 4 Remanent state MFM images of the lattices with scan size of 5×5 µm$^2$ (left side) and 2×2 µm$^2$ (right side). Solid circles and dotted circles indicate the positions of antidots. Dashed lines and dotted rectangular frames indicate the LE-SDWs and HE-SDWs, respectively.

FIG. 5 Different occurrences of HE-SDWs as a function of $S_x$, $S_y$ and $d$=60 (a), 80 (b) and 100 (c) nm. Shaded regions indicate the occurrence of HE-SDWs. Open circles indicate the lattices where no HE-SDWs are found; Half-filled circles (red part) indicate the lattices with narrow HE-SDWs; and filled circles (blue part) indicate the lattices with extended HE-SDWs.]. The right panels show two typical MFM images of the lattices with extended HE-SDWs and narrow HE-SDWs, corresponding to the filled (blue) and half-filled (red) circles, respectively.

FIG. 6 (a) Sketch of distribution of energy in an extended HE-SDW in the lattice with small $d$. (b) Change of energy sections in an extended HE-SDW and a narrow HE-SDW in the lattice with large $d$. The overall orientations of magnetization in different energy sections are indicated by black arrows.

FIG. 7 (a) and (b): two consecutive MFM images of the lattice with $d$=80 nm, $S_x$=350 nm, and $S_y$=300 nm. HE-SDWs that disappeared (1), whose widths were increased (2) and whose widths shrank (3) are indicated by arrows. (c): schematic diagram of the change of HE-SDW in the selected section marked by dotted rectangular in (b) during five consecutive MFM scans as indicated by different colors online. Cells that were occupied with sections of a HE-SDW during 26 scans are indicated by circled + signs. (d): histogram of the change in width (ΔW) of extended HE-SDWs (unchanged HE-SDWs are not included) from each scan during 26 consecutive MFM scans. The dashed line indicates a gaussian fitting curve.

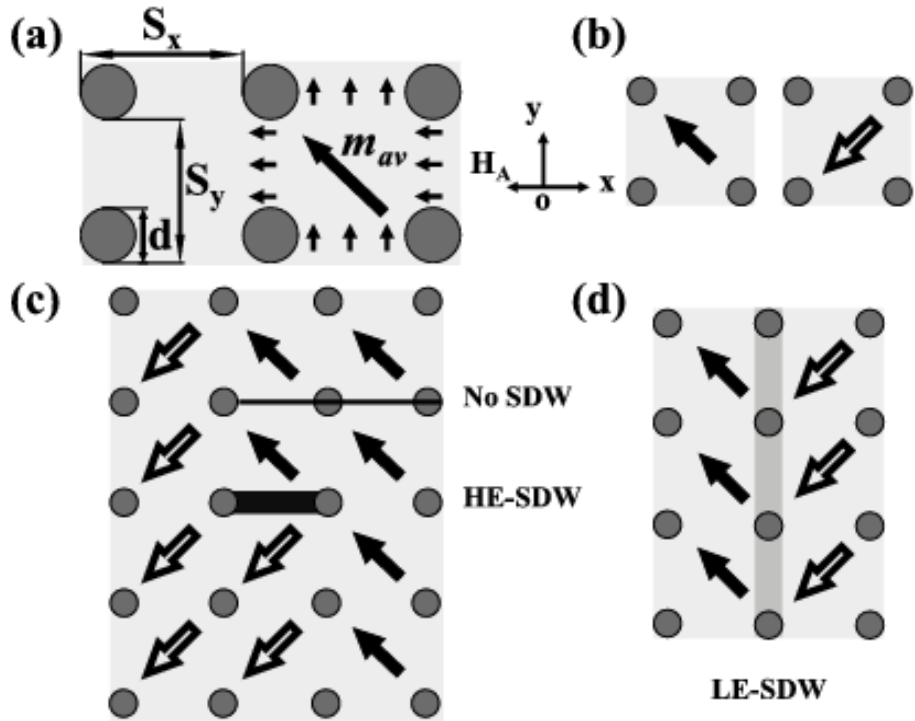

FIG. 1 (a) Schematic diagram of the permalloy antidot lattice and the domain distributions in one cell. (b) Two possible orientations of $m_{av}$ in experimental conditions. (c) HE-SDW in the $x$ direction. (d) LE-SDW in the $y$ direction.

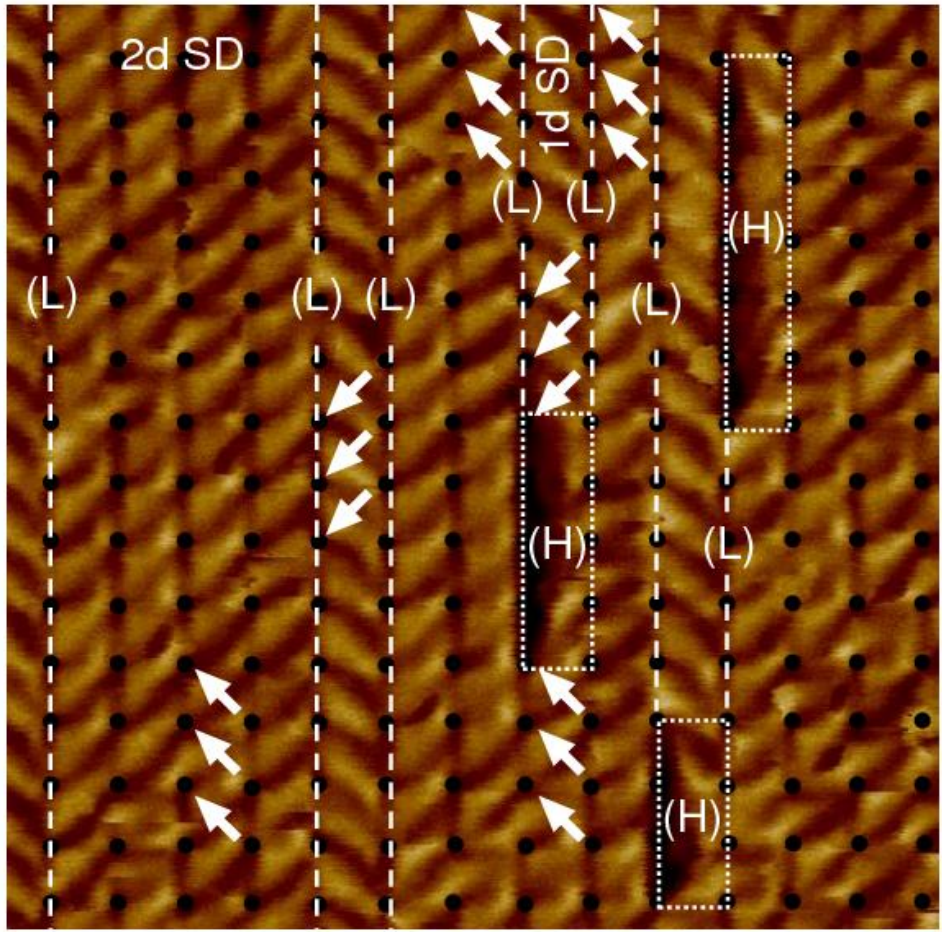

FIG. 2 Remanent state MFM images of the lattice: $d$=80 nm, $S_x$=350 nm and $S_y$=300 nm. Solid circles indicate the positions of antidots. Dashed lines and dotted rectangular frames indicate the LE-SDWs (L) and HE-SDWs (H), respectively.

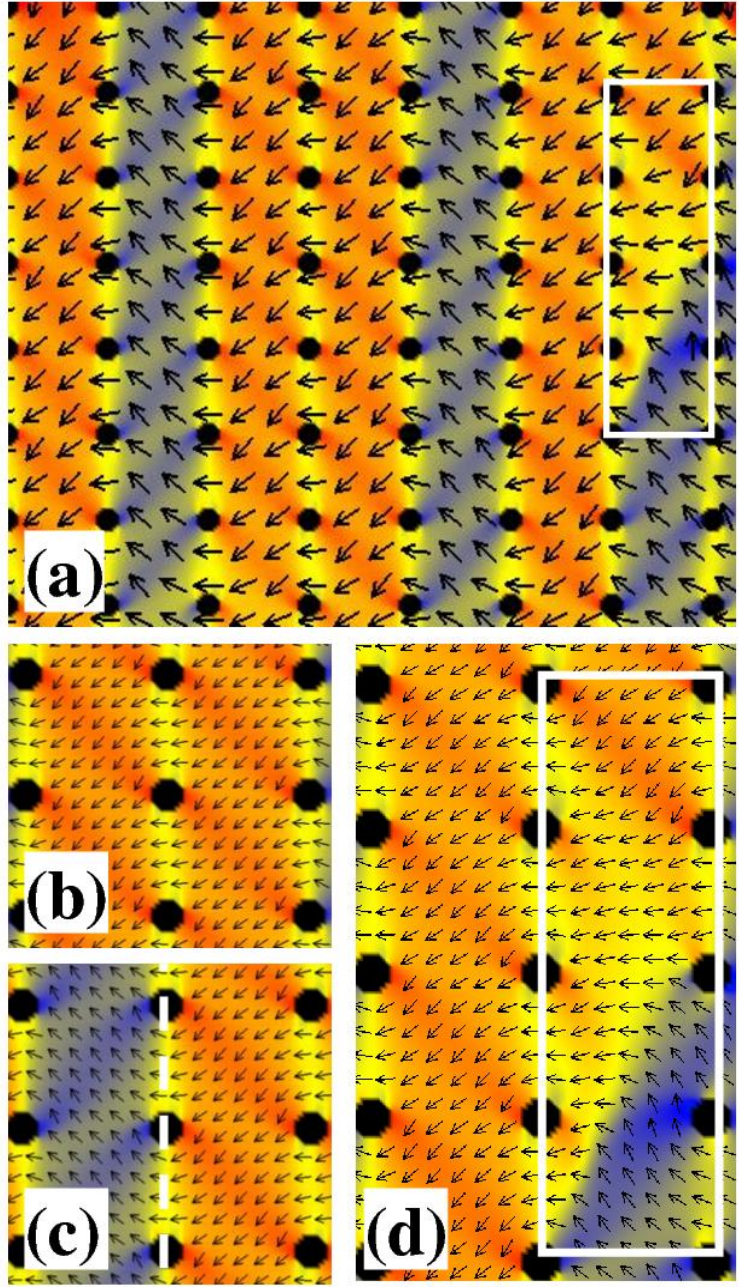

FIG. 3 The spin configurations of the lattice with $d$=80 nm, $S_x$=350 nm, and $S_y$=300 nm (a) and close-ups of SD (b), LE-SDW (c) and HE-SDW (d) obtained numerically. HE-SDWs are indicated by rectangular frames in (a) and (d). LE-SDW is indicated by the dashed line in (c).

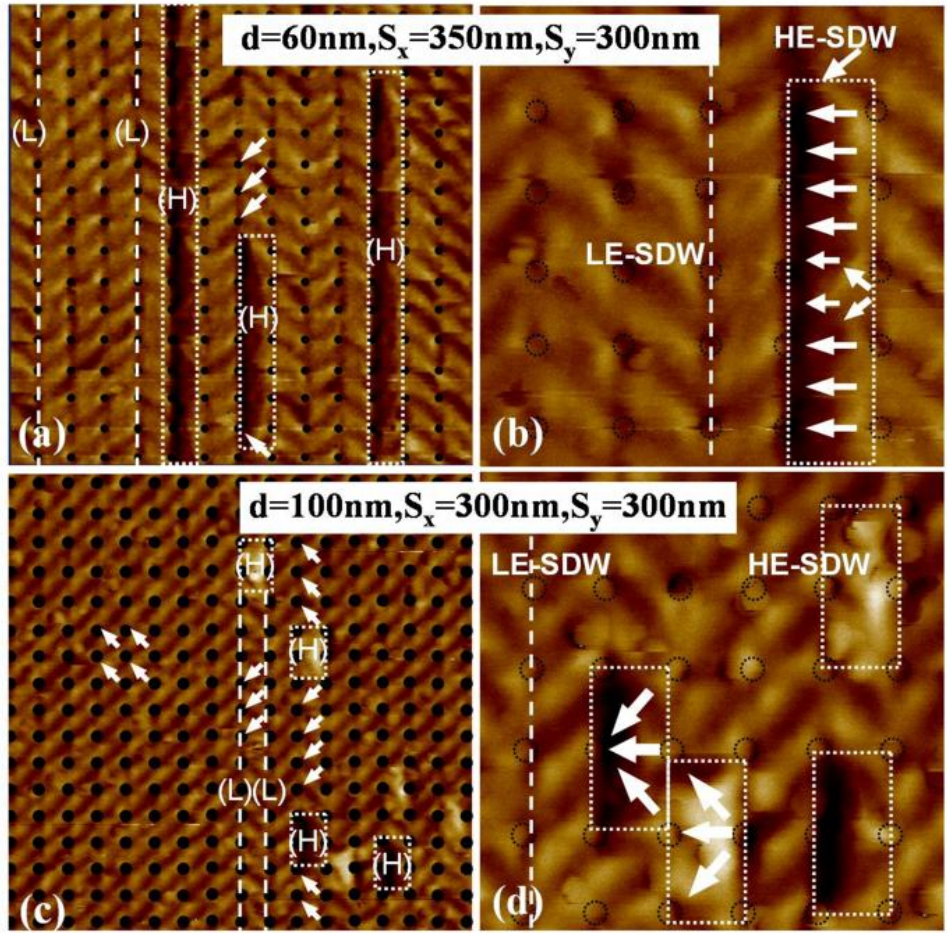

FIG. 4 Remanent state MFM images of the lattices with scan size of 5×5 μm² (left side) and 2×2 μm² (right side). Solid circles and dotted circles indicate the positions of antidots. Dashed lines and dotted rectangular frames indicate the LE-SDWs and HE-SDWs, respectively.

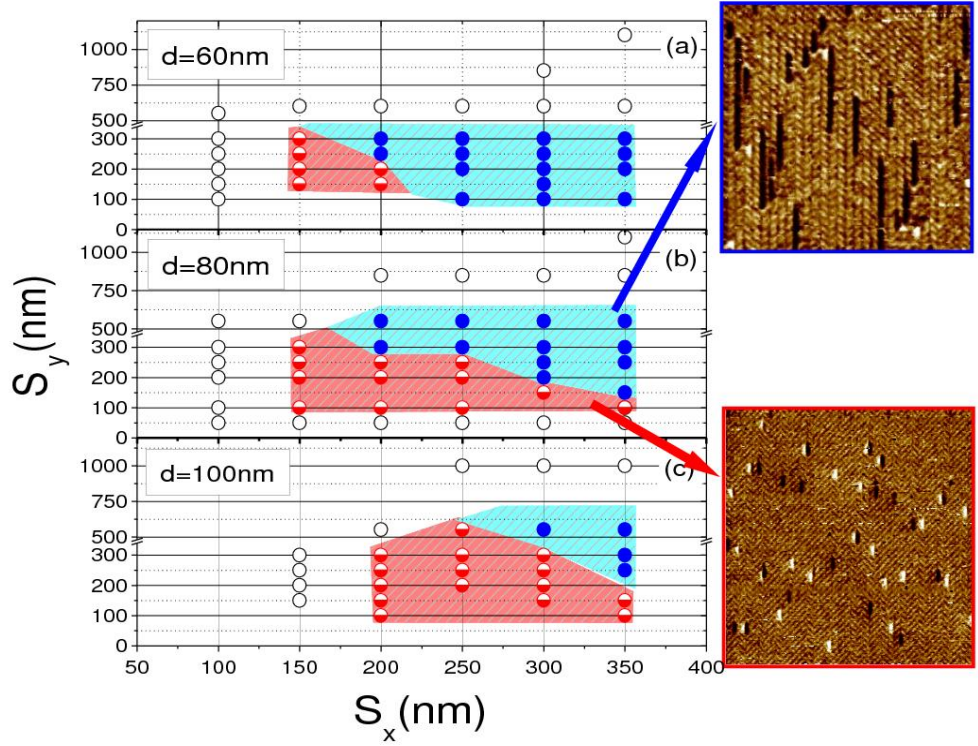

FIG. 5 Different occurrences of HE-SDWs as a function of $S_x$, $S_y$ and $d$=60 (a), 80 (b) and 100 (c) nm. Shaded regions indicate the occurrence of HE-SDWs. Open circles indicate the lattices where no HE-SDWs are found; Half-filled circles (red part) indicate the lattices with narrow HE-SDWs; and filled circles (blue part) indicate the lattices with extended HE-SDWs.]. The right panels show two typical MFM images of the lattices with extended HE-SDWs and narrow HE-SDWs, corresponding to the filled (blue) and half-filled (red) circles, respectively.

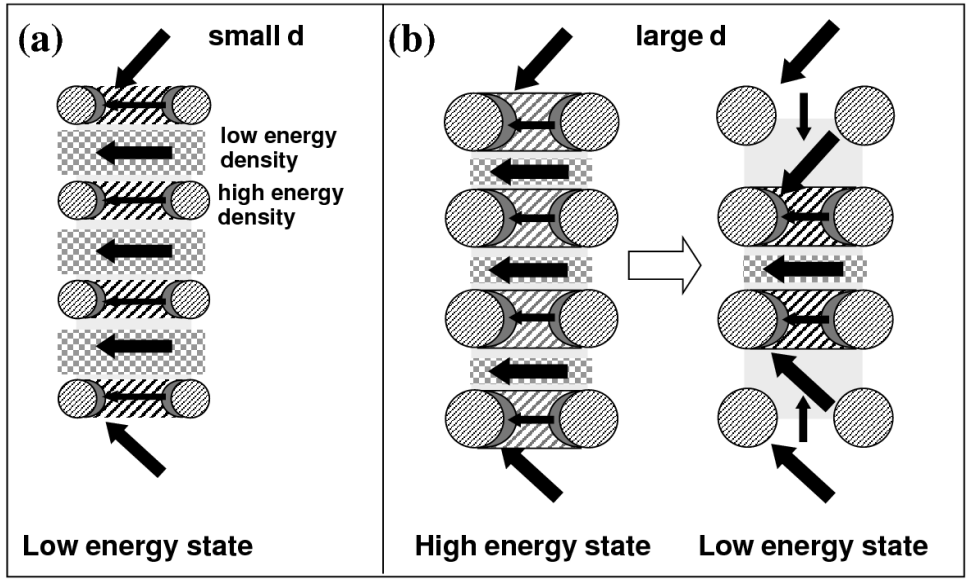

FIG. 6 (a) Sketch of distribution of energy in an extended HE-SDW in the lattice with small $d$. (b) Change of energy sections in an extended HE-SDW and a narrow HE-SDW in the lattice with large $d$. The overall orientations of magnetization in different energy sections are indicated by black arrows.

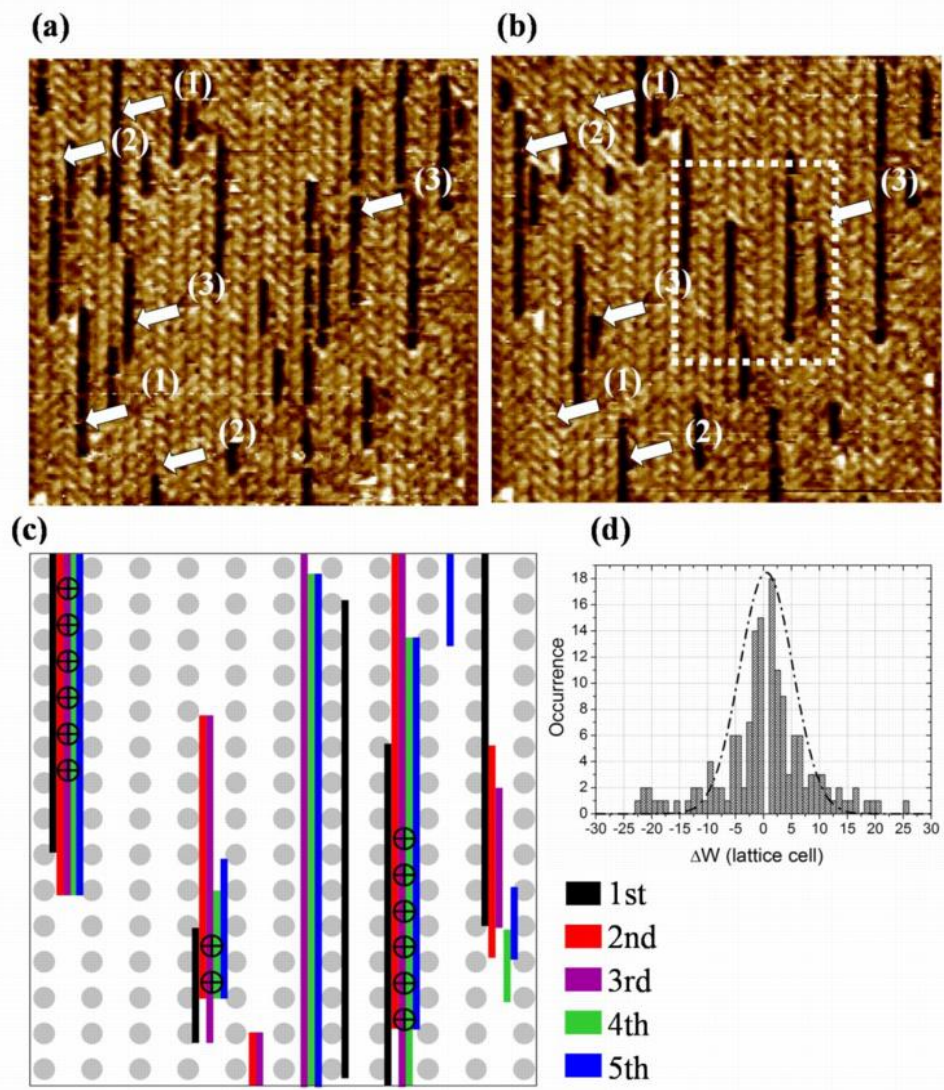

(a) and (b): two consecutive MFM images of the lattice with $d$=80 nm, $S_x$=350 nm, and $S_y$=300 nm. HE-SDWs that disappeared (1), whose widths were increased (2) and whose widths shrank (3) are indicated by arrows. (c): schematic diagram of the change of HE-SDW in the selected section marked by dotted rectangular in (b) during five consecutive MFM scans as indicated by different colors online. Cells that were occupied with sections of a HE-SDW during 26 scans are indicated by circled + signs. (d): histogram of the change in width ($\Delta W$) of extended HE-SDWs (unchanged HE-SDWs are not included) from each scan during 26 consecutive MFM scans. The dashed line indicates a gaussian fitting curve.